# Magnetoresistance and surface roughness study of the initial growth of electrodeposited Co/Cu multilayers

B.G. Tóth[+,*], L. Péter[#] and I. Bakonyi
Research Institute for Solid State Physics and Optics, Hungarian Academy of Sciences.
H-1525 Budapest, P.O.B. 49, Hungary
(August 18, 2011)

**Abstract** ─ The giant magnetoresistance (GMR) effect has been widely investigated on electrodeposited ferromagnetic/non-magnetic (FM/NM) multilayers generally containing a large number of bilayers. In most applications of the GMR effect, layered structures consisting of a relatively small number of consecutive FM and NM layers are used. It is of great interest, therefore, to investigate the initial stages of GMR multilayer film growth by electrodeposition. In the present work we have extended our previous studies on ED GMR multilayers to layered structures with a total thickness ranging from a few nanometers up to 70 nm. The evolution of the surface roughness and electrical transport properties of such ultrathin ED Co/Cu layered structures was investigated. Various layer combinations were produced including both Co and Cu either as starting or top layers in order (i) to see differences in the nucleation of the first layer and (ii) to trace out the effect of the so-called exchange reaction. Special attention was paid to measure the field-dependence of the magnetoresistance, MR(H) in order to derive information for the appearance of superparamagnetic regions in the magnetic layers. This proved to be helpful for monitoring the evolution of the layer microstructure at each step of the deposition sequence.

---

[+]Ph.D. student at Eötvös University, Budapest, Hungary
[*]Corresponding author. E-mail: tothb@szfki.hu
[#] Active member of The Electrochemical Society

**Introduction**

The giant magnetoresistance (GMR) effect in nanoscale ferromagnetic/non-magnetic (FM/NM) metallic multilayers has widespread applications today.[1,2]

Several different methods have been used for the preparation of such multilayers, among which electrodeposition is a simple but still sufficiently precise technique[3] to control the layer formation which is one of the main tasks for properly tuning the physical properties of the multilayers produced.

It was shown soon after the discovery of the GMR effect that electrodeposition is also capable of producing multilayer films with GMR.[4] The full literature of this field has been recently reviewed.[5] Various studies have been carried out to reveal how the deposition conditions (e.g., pH of the electrolyte[6], deposition pulse combination[7-9] or the non-magnetic layer deposition potential[8]) influence the GMR. Detailed structural investigations have also been performed to establish a correlation between multilayer microstructure and GMR.[7,9-12]

When looking at the relevant reports on electrodeposited (ED) FM/NM multilayer films with GMR behavior,[5] it turns out that most of the studies have been made on multilayers with a total thickness above 100 nm or even in the micrometer range. It has long been well-known, on the other hand, that films deposited with such large thicknesses usually undergo a significant roughening, regardless of the preparation method. This roughening effect was reviewed by Schwarzacher[13] for ED films of pure metals (such as Cu, Ag and Ni) and alloys (such as Fe-Co, Ni-Co and Ni-P). It was demonstrated for ED Co-Ni-Cu/Cu multilayer films[14] that such layered structures also show a continuous roughening up to a total film thickness of 1000 nm and beyond. Important findings of this latter work were that (i) on a rough substrate the multilayer film roughness was larger than on a smooth substrate although the dependence of roughness on total film thickness was much weaker for the first case and (ii) the roughness was much larger for a Cu deposition potential at which a significant dissolution of the magnetic layer is expected as compared to a potential with limited dissolution. Unfortunately, no magnetoresistance (MR) measurements were reported on these particular multilayer films.

Actually, the roughening at large total film thicknesses had been previously demonstrated for ED Co-Cu/Cu (Ref. 9) and Ni-Cu/Cu (Ref. 10) multilayers by cross-sectional transmission electron microscopy (TEM) which revealed a saw-tooth like top surface of the deposit. At the same time, large-magnification TEM could resolve a clear layered structure even at the top of the multilayer stack and a fairly large GMR was also detected on these multilayers.[9,10] It was, furthermore, established in another study[15] that the GMR magnitude was not significantly different for thick ED Co-Cu/Cu multilayer films when deposited either on mechanically polished, fairly rough Ti sheet substrates or on very smooth Si/Ta/Cu substrates with evaporated thin Ta and Cu underlayers. This may indicate that at large multilayer film thicknesses the overall microstructural features of the multilayers which are relevant for GMR become independent of the substrate. This conjecture is supported also by the finding that comparable GMR magnitudes were reported for thick ED Co/Cu multilayers obtained on polished Ti [16] and amorphous alloy ribbon [17] substrates.

The above mentioned roughness occurring on a large lateral scale (called also undulation) and causing a canting of the multilayer planes with respect to the film plane [10] may be beneficial for the GMR. Namely, in such a case the current flowing in the film plane crosses the plane of the individual layers, i.e., the GMR measured has also a "current-perpendicular-to-plane" (CPP) component. On the other hand, roughness with a small lateral length scale may be detrimental to the GMR due to the onset of an "orange-peel" coupling which is ferromagnetic in nature and leads to a reduction of the GMR as analyzed by Shima et al.[18] for ED Co/Cu multilayers. The small-scale interface roughness inherited from a rough substrate was demonstrated[19] to promote the appearance of superparamagnetic (SPM) regions in the magnetic layers which then give rise to a GMR contribution often not saturating even up to 10 kOe (Ref. 5). It was indeed reported that a rougher substrate leads to a larger SPM contribution to the GMR for Co/Cu multilayers produced by either electrodeposition[15] or sputtering.[19]

In most applications of the GMR effect, layered structures consisting of a relatively small number of consecutive FM and NM layers are generally used (so-called spin-valve structures[20]). It is of great interest, therefore, to investigate the initial stages of GMR multilayer film growth by electrodeposition.

However, there have been relatively few studies on the initial growth of ED GMR multilayers films close to the substrate. A cross-sectional TEM investigation on thick ED Co-Cu/Cu multilayers deposited on a Ti sheet substrate showed[21] that the first few bilayers are very disordered and a well-defined layered structure develops later only. This can be due to the fact that in this case the substrate surface is actually $TiO_2$, which does not provide a proper lattice matching for the nucleation and growth of multilayers with interest for GMR. Therefore, it is essential to use smooth substrates providing good lattice matching with the electrodeposited multilayer stack which can be ensured by using mostly an appropriate semiconductor wafer such as Si or GaAs, with or without a buffer layer. Attempts have been made along this line to prepare spin-valve like structures by electrodeposition[22-27] and fairly good spin-valve type GMR characteristics were obtained.

In an effort to grow similar thin layered structures with attractive GMR characteristics by electrodeposition and to better understand the initial stages of nucleation and growth of ED multilayers, our previous studies on thick GMR multilayers[7-12,15-17,21] have been extended to ultrathin ED Co/Cu layered structures. These ultrathin samples were deposited onto very smooth Si/Cr/Cu substrates with evaporated thin Cr and Cu underlayers providing a proper lattice matching with the growing layer structure. The evolution of the surface roughness and electrical transport properties (resistivity and magnetoresistance) of these samples was investigated. Various layer combinations were produced including both Co and Cu as starting or top layers in order (i) to see differences in the nucleation of the first layer and (ii) to trace out the effect of the so-called exchange reaction.[28-30] The latter implies dissolution of the Co atoms and their replacement by Cu atoms,[8,29] the whole process taking place with zero net current ($Co + Cu^{2+} \rightarrow Co^{2+} + Cu$). Special attention was paid to measure the field-dependence of the magnetoresistance, $MR(H)$, since it was shown previously[31] that from an analysis of the $MR(H)$ curves useful information can be derived for the SPM regions in the magnetic layers. This proved to be helpful for monitoring the evolution of the layer microstructure at each step of the deposition sequence.

**Experimental**

*Sample preparation.* — For the deposition of the Co/Cu multilayers, a sulfate/sulfamate type aqueous electrolyte was used. Its composition was 0.74 mol/ℓ $CoSO_4$, 0.010 mol/ℓ $CuSO_4$, 0.10 mol/ℓ $Na_2SO_4$, 0.25 mol/ℓ $H_3BO_3$ and 0.25 mol/ℓ $HSO_3NH_2$. The pH was set to 3.25 by adding NaOH to the solution. The choice of this pH value was based on some preliminary experiments to get appropriate deposition conditions.

The Co/Cu multilayers were deposited on a [100]-oriented, 0.26 mm thick Si wafer covered with a 5 nm Cr and a 20 nm Cu layer by evaporation. The purpose of the chromium layer was to assure adhesion and the Cu layer was used to provide the electrical conductivity of the cathode surface.

The deposition was performed by pulse plating in a tubular cell of 8 mm x 20 mm cross section with an upward looking cathode at the bottom of the cell.[32] For the deposition of the magnetic layer, galvanostatic (G) control mode was used at -35.1 mA/cm$^2$ current density. At these high current densities, less than 1 at.% Cu gets incorporated in the magnetic layer, which does not deteriorate the magnetic and transport properties of the layer. For the Cu layer, potentiostatic (P) control mode was used at a potential of -0.585 V vs. the saturated calomel electrode (SCE). According to a previous optimization, at this Cu deposition potential neither dissolution of the Co layer, nor Co codeposition into the Cu layer can occur.[33] This also ensures that when controlling the layer thicknesses by adjusting the deposition time in the G mode and by setting up the charge passed through the cell in the P mode, the resulting layer thicknesses will be close to the predetermined values. For bulk Ni-Co layers, previous profilometric measurements[34] established that the current efficiency is high enough, namely 96 %, to assume that the actual layer thicknesses are fairly close to the preset values. Furthermore, XRD and TEM studies[11,12] indicated that under such controlled multilayer deposition conditions, the actual layer thicknesses are, indeed, only slightly above the nominal values.

The deposition pulse sequence was manipulated in various manners. For series 1, Co(2.0 nm) and Cu(5.0 nm) layers were alternately deposited onto each other by increasing the total number of individual layers by one for each subsequent sample until the formation of a stack of 5 Co/Cu bilayers after which the increment was one Co/Cu pair up to 10 Co/Cu bilayers. The choice of the individual layer thicknesses was based on our previous experience[16,35] obtained on ED GMR Co/Cu multilayers prepared with optimized Cu deposition potential. This deposition sequence resulted in samples with either a Co or a Cu terminal layer on the top of the multilayer sample. The aim of this series was to investigate the variation of surface roughening and magnetoresistance layer by layer. A sketch of the substrate and deposit structure for series 1 is shown in Fig. 1a.

In order to investigate how the exchange reaction between the covering Co layer and the electrolyte affects the resulting GMR in series 1, four additional samples (series T) were made with the following structure: Co(2.0 nm)/T/Cu(5.0 nm)/Co(2.0 nm)/T/Cu(5.0 nm). In this pulse sequence, $I = 0$ mA was set for 0, 5, 10 and 20 seconds during the "T" pulse (what is equivalent to the open circuit conditions).

Another set of multilayers (series 2) was also prepared for which the deposition process consisted of the repetitions of a trilayer sequence Cu(2.5 nm)/Co(2.0 nm)/Cu(2.5 nm). This way, the Co layers were protected from the exchange reaction at each stage of the multilayer formation. The individual layer thicknesses were controlled in a manner that, apart from the insertion of the very first Cu layer in series 2, identical multilayer stacks were obtained in the two series. This is because we can think of the multilayers in series 2 (sequence [Cu/Co/Cu]$_N$) as Cu(2.5 nm)/Co(2.0 nm)/[Cu(5 nm)/Co(2.0 nm)]$_{N-1}$/Cu(2.5 nm) where n is the number of the deposited trilayers (see sketch in Fig. 1b). The additional Cu layer deposited directly on the Si/Cr/Cu substrate had two important consequences. Firstly, it provided a fresh, non-oxidized Cu surface also for the deposition of the very first Co layer. Secondly, the deposition of the starting Cu layer in series 2 depleted the electrolyte for the Cu$^{2+}$ ions immediately at the cathode surface by the time the Co deposition pulse started. Therefore, a different amount of Cu codeposited with Co in the first magnetic layer can be expected for the two series what may have an influence on the observed magnetoresistance.

*Measurement of surface roughness and microstructure.* —The root-mean-square surface roughness ($R_q$) of the multilayers was investigated with atomic force microscopy (AFM). The surface roughness of the Si/Cr/Cu substrate with the evaporated underlayers was also determined and was found to show height fluctuations not larger than 3 nm.

For one of the thickest multilayer stack (70 nm), we have carried out an X-ray diffraction study which revealed a very dominant (111) texture. This is in agreement with our previous experience according to which electrodeposited Co/Cu multilayers prepared under similar conditions develop a (111) texture.[9,11,12,36]

*Measurement and evaluation of electrical transport properties.* —The zero-field resistivity ($\rho_0$) was determined at room-temperature with a four-point-in-line probe calibrated with Cu-foils of known thickness and having the same lateral dimensions as the alloy sample.

The magnetoresistance was measured with another four-point-in-line probe as a function of the external magnetic field ($H$) up to 8 kOe. The MR ratio was defined with the formula

$$MR(H) = \frac{\Delta R(H)}{R_0} = \frac{R(H) - R_0}{R_0} \qquad (1)$$

where $R_0$ is the resistance of the sample in zero external magnetic field and $R(H)$ is the resistance in an external magnetic field $H$. The magnetoresistance was determined in the field-in-plane/current-in-plane geometry in both the longitudinal (LMR, magnetic field parallel to the current) and the transverse (TMR, field perpendicular to the current) configurations. The measured $MR(H)$ curves were decomposed into GMR$_{FM}$ and GMR$_{SPM}$ contributions according to a procedure described previously.[31]

The MR data were measured on the multilayers while being on their substrates, which necessitates a correction for the shunt effect of the underlayers. This can be done by using the measured values of the zero-field resistivity $\rho_0$ of both the substrate alone and the substrate/multilayer stack.

## Results and Discussion

The major objective of the present work was to investigate the formation of ultrathin ED Co/Cu multilayers. For this purpose, the structure evolution with total deposit thickness was compared for series 1 and series 2 on the basis of the results of surface roughness and electrical transport (resistivity and magnetoresistance) measurements. The main differences of the two series occurred in the starting layer (Co in series 1 and Cu in series 2) on the evaporated underlayer and in the top layer of the stack (Co or Cu alternately in series 1 and always Cu in series 2). For both series, the $MR(H)$ curves are analyzed first for the initial layer stacks, followed by a discussion of the evolution of GMR with total multilayer thickness and, finally, the surface roughness data are presented and discussed.

*MR(H) curves of the initial layer stacks in series 1.* —Figure 2 shows the $MR(H)$ curves for the first four layer stacks in series 1: (a) the first Co layer electrodeposited on the Si/Cr/Cu substrate; (b) the first Co/Cu bilayer; (c) the trilayer Co/Cu/Co and (d) the first two bilayers Co/Cu/Co/Cu.

The first Co layer on the Si/Cr/Cu substrate (Fig. 2a) shows very weak anisotropic magnetoresistance (AMR) behavior ($LMR > 0$, $TMR < 0$) as expected for a single magnetic layer. The splitting of the $LMR(H)$ and $TMR(H)$ curves is due to spin-dependent scattering events in the bulk of the magnetic layers.[34,37,38] The magnitude of the AMR is defined as the difference $LMR - TMR$ at the highest magnetic field used. The very small value of the *AMR* magnitude (ca. 0.005 %) of the Si/Cr/Cu//Co sample (a single uncovered Co layer on the top) is due to the removal of most of the magnetic layer due to the exchange reaction.[29] Another possible factor leading to a reduction of the amount of magnetic material in this first magnetic layer is that during its deposition the $Cu^{2+}$ content of the bath at the cathode/electrolyte interface is still close to the bulk value. This means that Cu is deposited at much larger current density than the diffusion-limited current density. Thus, the amount of Cu codeposited with Co in the very first magnetic layer can be fairly large. We should also take into account that although under the deposition conditions of the magnetic layer the native oxide scale of the evaporated Cu layer is reduced to a large extent to metallic Cu, this reduction is certainly not complete over the whole cathode area and there may be local areas where the nucleation of the first Co layer may be strongly hindered. As a result, the first magnetic layer is definitely discontinuous; it may contain regions of high Cu content (or even pure Cu regions). By taking into account all these factors and the magnetoresistance results, the cross-sectional view of the Si/Cr/Cu//Co sample can be visualized as shown in Fig. 3a.

Putting a covering Cu layer on this first Co layer changes significantly the situation as revealed by the $MR(H)$ curves in Fig. 2b. By the exclusion of the exchange reaction, all the Co atoms deposited with a thickness of 2 nm are retained and the AMR magnitude increases to about 0.015 %, i.e., by about a factor of 3. Furthermore, in spite of the fact that we have to deal with a single magnetic layer only in this Si/Cr/Cu//Co/Cu stack, the observed magnetoresistance contains, surprisingly, a GMR contribution as well (both $LMR$ and $TMR$ are negative for all fields) which is comparable in magnitude to the AMR. The cross-sectional sketch of this sample in Fig. 3b may help in understanding the occurrence of the observed GMR effect. Due to the laterally non-continuous nucleation of Co on the evaporated Cu and the presence of a fairly high Cu content in the first magnetic layer which presumably forms segregations, the Co layer having a nominal total thickness of 2 nm is not completely homogeneous but may consist of separated Co islands. In case the magnetizations of the adjacent islands are not aligned parallel, polarized electrons travelling between two such adjacent islands may contribute to a GMR effect. These pathways may have a non-negligible probability since (i) there is now a conducting spacer material (Cu) both below and above the magnetic layer and, perhaps even more importantly, (ii) the discontinuities of the first magnetic layer are filled up by the covering Cu layer which provides a good conducting path for electrons between adjacent but separated ferromagnetic Co regions within the magnetic layer plane itself (see sketch in Fig. 3b). Due to the data scatter of the $MR(H)$ curves in Fig. 2b, we cannot ascertain either the presence or absence of an SPM contribution to the magnetoresistance curve.

The $MR(H)$ curves of the layer stack Si/Cr/Cu//Co/Cu/Co (Fig. 2c) indicate a further increase of both the AMR and GMR magnitudes by nearly a factor of two due to the presence of a second, albeit strongly dissolved Co layer on the top of the sequence. A Langevin fit could be carried out here for the transverse component and the decomposition showed that the $GMR_{FM}$ term is about -0.030 % whereas the $GMR_{SPM}$ term is about -0.022 %, i.e., the two contributions are of comparable magnitude. This indicates that the total GMR comes from several different contributions. The first is the electron scattering between Co regions in the first Co layer (either FM or SPM). Another contribution comes from the scattering between large ferromagnetic Co regions in the first layer and ferromagnetic Co islands in the second Co layer ($GMR_{FM}$ term). In addition, there are frequent electronic transitions also between FM regions of one layer and SPM regions of the other layer ($GMR_{SPM}$ term) as sketched in Fig. 3c.

The *MR*(*H*) curves for the next multilayer stack Si/Cr/Cu//Co/Cu/Co/Cu are shown in Fig. 2d. The layer sequence applied ensures the suppression of the exchange reaction for both Co layers and this enables the formation of well-defined ferromagnetic layers which are properly separated from each other by the first Cu layer (see the schematic view of the cross-section of this layer stack in Fig. 3d). This is confirmed by the observed *MR*(*H*) curve which indicates a clear GMR behavior due to electron spin scattering events for electrons travelling between the two magnetic layers through the non-magnetic spacer layer. The GMR is dominated by a ferromagnetic contribution $GMR_{FM}$ (ca. -0.47(2) %) characterized by a low saturation field (below 1 kOe). This indicates that the magnetic layers consist mainly of ferromagnetic regions and the fraction of SPM regions in the magnetic layers is not large (the SPM contribution $GMR_{SPM}$ which derives from electrons travelling between a FM and an SPM entity is very small, ca. -0.05(1) %). The SPM regions can form in the first magnetic layer due to its higher Cu content via phase separation[11,12] or in both magnetic layers through a mechanism induced by small-lateral scale surface roughness as suggested by Ishiji and Hashizume.[19] It should also be noted that the GMR effect of the stack is by an order of magnitude larger than obtained for the previous stack Si/Cr/Cu//Co/Cu/Co (Fig. 2c).

*Influence of exchange reaction on magnetoresistance.* —In order to underpin the crucial role of the exchange reaction the presence of which was utilized in the explanation of some of the *MR*(*H*) in the previous section, we shall now present results for series T specially designed for this purpose. The starting point of the series Co(2.0 nm)/T/Cu(5.0 nm)/Co(2.0 nm)/T/Cu(5.0 nm) was the layer stack prepared with *T* = 0 s which sample was identical with the multilayer Co/Cu/Co/Cu of series 1 (see Fig. 2d). In this layer stack, both magnetic layers are immediately covered by an ED Cu layer in order to prevent the occurrence of an exchange reaction on the Co layers.

As soon as the T pause time has a non-zero value for the series Co(2.0 nm)/T/Cu(5.0 nm)/Co(2.0 nm)/T/Cu(5.0 nm), a spontaneous exchange reaction starts between the Co atoms in the lastly deposited magnetic layer and the $Cu^{2+}$ ions in the solution. As the T pause time increases more and more Co atoms are dissolved into the electrolyte and substituted with Cu due to the exchange reaction. This finally leads to a reduction of the magnetic layer thickness and, at the extreme, to a fragmentation of the magnetic layer which, therefore, becomes discontinuous. As a result of the exchange process, extended regions of the magnetic layer may be sufficiently large to exhibit FM behavior whereas smaller and magnetically separated Co islands may show SPM behavior.

These effects can clearly be seen in Fig. 4 as the T pause time increases from 0 to 20 s. Not only the absolute value of the GMR decreases abruptly already at 5 s pause time but also the relative ratio of the SPM component in the total magnetoresistance increases gradually with increasing pause time. Whereas the first sample in Fig. 4a shows a clear GMR effect with mainly FM contribution, for the last one (Fig. 4d) the very weak GMR is due to the dominant FM-SPM contribution. The last sample also exhibits a very small AMR effect, indicating that the exchange reaction has reduced the lateral size of most continuous Co segments to an extent that in these small magnetic entities two consecutive spin-dependent scattering events can hardly occur anymore.

*Evolution of the magnetoresistance with total multilayer thickness for series 1.* —The rest of the multilayers in series 1 which were obtained by the addition of more and more Co/Cu bilayers exhibited *MR*(*H*) curves very similar to that seen in Fig. 2d, with a further increase of the GMR magnitude as the number of bilayers increased.

Figure 5 shows the measured $GMR_{FM}$ components of the multilayers of series 1 (open symbols) and also the data obtained after correcting for the shunting effect of the substrate by using the measured resistivity (4.6 μΩ cm) of the Si/Cr/Cu substrate (closed symbols).

The overall evolution of the corrected $GMR_{FM}$ component with total layer thickness appears as a monotonous increase up to about 40 nm and then it remains nearly constant. However, this initial increase is not at all monotonous if we consider incremental steps of adding another layer to the previous stack. In the thickness range below 40 nm where the thickness increment was either a single Co or a single Cu layer, the influence of the addition of a subsequent layer was dependent on whether it was Co or Cu. It can be observed that the addition of a Cu layer increased the GMR whereas the addition of a Co layer hardly influenced it. The explanation lies again in the exchange reaction. When a Co layer is put on the top, it is mostly removed by the exchange reaction by the time its surface is made free of the electrolyte and, therefore, it hardly offers new regions for a magnetic/non-magnetic/magnetic sequence which is a pre-requisite for the occurrence of a GMR effect. On the other hand, when the top layer is Cu, it effectively protects the underlying last Co layer from the effect of the exchange reaction. Therefore, the last Co layer, together with the previous Co layer and the Cu layer in between, can form a good magnetic/non-magnetic/magnetic sequence effectively contributing to the GMR. The increase of the GMR effect continues until the new subsequent layers effectively improve the multilayer structure in the sense that more and more areas provide contributions to the GMR via forming proper magnetic/non-magnetic/magnetic layer sequences. With reference to Fig. 5, this seems to be the case up to a total thickness of about 40 nm.

The $GMR_{SPM}$ term showed a similar evolution with multilayer thickness as the $GMR_{FM}$ contribution. However, as can be inferred from the shape of $MR(H)$ curve for the multilayer stack Si/Cr/Cu//Co/Cu/Co/Cu (Fig. 2d), its magnitude is much smaller than the $GMR_{FM}$ term (the relative fraction of $GMR_{SPM}$ contribution to the total observed magnetoresistance remained below about 0.1 for the whole series 1).

*Evolution of surface roughness for series 1.* —The root-mean-square surface roughness for multilayer series 1 varied apparently randomly in a wide range with the addition of new layers to the stack (Fig. 6). The first Co layer subjected to the exchange reaction exhibited a large increase of the surface roughness with respect to the substrate roughness. This may be partly attributed to the hindered and, therefore, uneven nucleation of Co on the surface of the evaporated Cu underlayer from which the native oxide was not removed. Another possible reason of the surface roughness increment was the pronounced dissolution of the Co atoms and their random replacement by Cu atoms. The deposition of the next Cu layer leads to a smoothening. Since the second and subsequent Co layers are already deposited on a completely oxide-free surface, the nucleation of Co is much more homogeneous over the cathode area. Therefore, the top Co layers, in spite of the exchange reaction, lead to a smoothening, at least up to the fifth bilayer as indicated by the lines connecting the full circles (top Cu layer) with the subsequent open circles (top Co layer). On the contrary, the top Cu layers rather seem to cause a roughening in most cases. In summary, it seems that the dissolution makes the multilayer smoother: the samples with the partially dissolved Co covering layer are always smoother than the ones with a Cu layer on the top.

To prove this smoothening effect of the substitution of Co atoms by Cu atoms, a series of four samples was made. Each was made up by four bilayers of Cu(5.0 nm)/Co(2.0 nm) and, after finishing the deposition, the current was set to zero but the electrolyte was not removed from the sample for different waiting times ($T_W$). During this period, a replacement of the Co atoms with Cu atoms by the exchange reaction could occur. The root-mean-square surface roughness of the resulting sample was measured by AFM. According to Fig. 7, with increasing $T_W$ the $R_q$ values decrease exponentially. This is because both partial processes of the exchange reaction result in a roughness reduction. On the one hand, the dissolution of the Co atoms at highest peaks has the largest probability because the binding energy of these atoms to the solid phase is the smallest. On the other hand, the deposition of the Cu atoms and their diffusion along the surface may lead to a filling-up of the cavities because the binding energy of the newly deposited atoms at these positions is the highest. The $R_q$ values converge to a finite value (4.2 nm) which is somewhat higher than the initial substrate roughness. One should keep in mind that the lattice mismatch in a multilayer should unavoidably lead to a surface roughening because the layer with larger lattice constant can grow in island like form only at the beginning on the previous layer with smaller lattice constant.

In fact, the exponential decrease of the $R_q$ values starts only after a certain waiting time (which is lower than the minimum 10 s we could reach), because the electrolyte near the surface is depleted for $Cu^{2+}$ ions and thus the exchange reaction cannot start immediately after the deposition of the Co layer was finished. We can assume that the roughness starts to decrease at a very low rate, due to the lack of $Cu^{2+}$ ions, and then, as the $Cu^{2+}$-content of the electrolyte near the sample surface reaches its bulk value, the smoothening rate increases. After a certain value of $T_W$, the amount of the Co atoms at the surface decreases to a low value as a result of which the $R_q$ data converge to a finite value.

The contribution of this smoothening effect to the total surface roughness measured in the case of the first few samples in series 1 changes randomly because the $T_W$ data of these samples were not measured (albeit it was close to the minimum electrolyte removal time of 10 s that we could reach).

*MR(H) curves of the initial layer stacks in series 2.* —Figure 8 shows the *MR(H)* curves for the first two layer stacks in series 2: (a) Si/Cr/Cu//[Cu2.5nm)/Co(2.0nm)/Cu(2.5nm)]x1 and (b) Si/Cr/Cu//[Cu2.5nm)/Co(2.0nm)/Cu(2.5nm)]x2. The ED layer stack in Fig. 8a is actually a Cu/Co/Cu sequence. The magnetic layer in this sequence is expected to consist mainly of FM regions since its Cu content is low and the top Cu layer prevented it from the exchange reaction. Correspondingly, a bulk AMR effect of the FM Co layer is expected to occur as shown, indeed, by the *MR(H)* curves (*LMR* > 0 and *TMR* < 0) in Fig. 8a. The magnitude of the AMR effect (the splitting of the LMR and TMR curves in the saturation region) is larger than was for the first single Co layer of series 1 (stack Si/Cr/Cu//Co, see Fig. 2a). This is because of the larger amount of the magnetic material for the first magnetic layer in series 2. There is also a difference in comparison with the layer stack Si/Cr/Cu//Co/Cu of series 1 (see Fig. 2b). The latter one also contains a layer stack Cu//Co/Cu but the discontinuity of the magnetic layer as discussed in previous sections for this layer stack could give rise to a GMR effect. However, the Co layer in the Cu//Cu/Co/Cu stack is not expected to show GMR due to its continuous layer form. Therefore, this single Co layer behaves as a bulk magnetic material.

According to the $MR(H)$ curves in Fig. 8b, the second layer stack Si/Cr/Cu//Cu/Co/Cu/Cu/Co/Cu of series 2 already exhibits a clear GMR behavior with a dominant FM contribution. This is because it contains the sequence Co/Cu/Co which is, apart from the microstructure of the first magnetic layer, identical with the similar sequence in the layer stack of Si/Cr/Cu//Co/Cu/Co/Cu of series 1 (see Fig. 2d). The larger GMR magnitude of the multilayer in series 1 may come from the discontinuous nature of the first magnetic layer in this stack which can provide electron pathways for GMR also in the layer plane between separated FM regions and not only electron pathways between the two magnetic layers which is the case for the stack of series 2.

The rest of the multilayers in series 2 exhibited $MR(H)$ curves qualitatively very similar to those shown in Fig. 8b for the stack consisting of two Cu/Co/Cu trilayers, just the GMR magnitude varied with the number of trilayers as will be discussed in the next section.

*Evolution of the magnetoresistance with total multilayer thickness for series 2.* —In order to evaluate the evolution of the magnetoresistance for series 2, first we consider the zero-field resistivity data. In contrast to series 1, the resistivity data for multilayers of series 2 on their substrates showed a systematic variation with total multilayer thickness (Fig. 9) which was described according to the following considerations. The stack of the substrate and the multilayer can be treated as two resistors in a parallel configuration with resistivities $\rho_S$ and $\rho_{ML}$ and thicknesses $d_S$ and $d_{ML}$ where the subscripts S and ML refer to the substrate and the multilayer, respectively. The measured resistivity ($\rho_0$) can be described as

$$\rho_0 = \rho_{ML} \frac{1 + \dfrac{d_{ML}}{d_S}}{\dfrac{\rho_{ML}}{\rho_S} + \dfrac{d_{ML}}{d_S}} \qquad (1)$$

In deriving eq. (1), it was assumed that the individual layers of the multilayers are perfectly flat and smooth. This expression was fitted to the experimental data in Fig. 9 in which the multilayer resistivity $\rho_{ML}$ is the only fitting parameter assumed to be constant throughout the series. The fitted curve also shown in Fig. 9 describes the data fairly well. A consistency of the fit is indicated by displaying the fitted constant $\rho_{ML}$ value (6.6 μΩ cm) as a horizontal line corresponding well to the average of the shunt-corrected resistivity values of the multilayers throughout multilayer series 2. By using these latter values, we can correct the measured magnetoresistance data for the shunting effect of the substrate.

The measured and corrected $MR_{FM}$ data are displayed in Fig. 10. After the third trilayer stack, a sudden increase of the $GMR_{FM}$ term can be seen and, then, a leveling off roughly for the same total multilayer thicknesses as for series 1. The saturation $GMR_{FM}$ value is somewhat larger for series 1 but the origin of this difference is not clear.

The $GMR_{SPM}$ contribution for series 2 showed a monotonous, nearly linear increase with trilayer number (Fig. 10). Its relative fraction in the total magnetoresistance was somewhat higher (varied between 0.1 and 0.18) than for series 1, although it remained also here much less than the $GMR_{FM}$ term.

*Evolution of surface roughness for series 2.* —For multilayer series 2 (sequence Si/Cr/Cu//[Cu/Co/Cu]$_N$), we obtained a much more systematic change in the $R_q$ parameter (Fig. 11) because of the more controlled deposition and dissolution conditions. Even up to 70 nm total thickness, an approximately linear increase of the surface roughness was detected. Representative AFM pictures of the Si/Cr/Cu substrate and the thickest multilayer in this series are shown in Fig. 12.

Contrary to the alternating deposition of Co and Cu from sample to sample in series 1, in the second series the same trilayer structure was deposited repeatedly. This led to a homogeneous structural evolution of the deposited multilayers, the only difference between the subsequent samples being the number of trilayers. This allows us to compare multilayers in series 2 on the basis of their total thickness only.

Since the first and the last component of the trilayer are both Cu, they behave as a continuously deposited 5 nm thick layer in the multilayer stack. This means that the roughening of the multilayer cannot come from the trilayer-trilayer interface because Cu is continuously deposited there. Therefore, an increase of the $R_q$ parameter with thickness derives partly from the lattice mismatch (and thus the island-like growth) between the Co and the Cu layers as noticed above and partly from the normal cumulative roughening of the individual layers during their growth.

Apart from the linear surface roughness growth for multilayers presented in this paper other authors have reported different roughening properties of alloys and multilayers. For metal layers (specifically Cu) a linear increase of the root-mean-square roughness was found by Schwarzacher[13] within the range of 0.2 and 20 µm. Renner and Liddell[39] reported the saturation of $R_q$, also for Cu, between the thicknesses 1 and 12 nm. For Ni-Co/Cu multilayers in the 8 to 800 nm thickness range, da Silva and Schwarzacher[14] found an exponential roughening. These results along with the roughness data given in this paper underline the necessity of further systematic studies of the effect of substrate and even the electrolyte on the surface properties of the deposited alloys and multilayers.

For large total thicknesses, the roughness in series 2 exceeds by about a factor of two the value of the roughness for most of the multilayers in series 1. This may be partly an explanation for the lower saturation $GMR_{FM}$ values obtained for series 2. Also, the linear increase of the $GMR_{SPM}$ term and the roughness for series 2 (Fig. 10) are so well correlated that this gives strong support for the validity of the model put forward by Ishiji and Hashizume[19] for the origin of the formation of SPM regions in FM/NM multilayers.

*Microstructure evolution in early stages of multilayer formation*. —By comparing the MR data of the first sample of series 1 (showing a very small AMR) and series 2 (showing a clear AMR), it can be established that depositing Co on a Cu layer of the evaporated substrate with the probable presence of residual oxide regions makes the Co atoms form islands, either FM and SPM-type. If this first Co layer is covered with a Cu layer, the "bilayer" shows GMR, which is a sign of electron scattering between these islands. On the contrary, if the evaporated substrate Cu layer is first covered with an electrodeposited, fresh Cu layer, the Co atoms form an almost continuous, predominantly ferromagnetic layer. If this is covered with a subsequent ED Cu layer in order to prevent the exchange reaction, the sample shows a good AMR behavior as a sign of a mostly continuous layer of Co.

In the third sample of series 1 (Co/Cu/Co), scattering occurs between smaller (SPM) and larger (FM) islands of Co below and above the electrodeposited Cu layer because of the dissolution of the top Co layer. However, if an additional protective Cu layer is deposited, electron scattering occurs predominantly between large FM regions of the two layers making the multilayer to show a definite GMR. The same behavior with almost the same MR value can be seen for the second sample in series 2, which also contains two Co layers with a protective topmost Cu layer.

If more than two Co and Cu layers are deposited in series 1, the MR behavior remains the same because the electron scattering between these almost continuous Co layers dominate.

**Summary**

In the present work, the evolution of the microstructure of electrodeposited Co/Cu multilayers was investigated. For this purpose, two different Co/Cu multilayer series were prepared: one with alternating Co and Cu layers on the top and another one made up of Cu/Co/Cu trilayers on Si wafers with evaporated Cr and Cu underlayers. The thicknesses of the multilayers were varied between 7 and 70 nm.

It was shown by AFM studies that although the exchange reaction between a topmost Co layer and the electrolyte (series 1) deteriorates the continuity of the top Co layer and, thus, leads to the appearance of an SPM contribution to the magnetoresistance, it also makes the multilayer smoother.

It was also found that, in the multilayer thickness range investigated, the root-mean-square roughness of the multilayers made up by the trilayers (series 2) develops linearly as the total multilayer thickness increases. Meanwhile, the GMR of the deposited multilayer saturates at six trilayers whereas the small SPM contribution to the MR increases in the whole investigated thickness range.

**Acknowledgements**

This work was supported by the Hungarian Scientific Research Fund through grant OTKA K 75008. The authors acknowledge G. Molnár (Research Institute for Technical Physics and Materials Science, HAS) for preparing the evaporated underlayers, Á. Pekker for his help in the handling of the AFM equipment and G. Faigel for making the AFM available for the present research. We also acknowledge L.K. Varga for the XRD study.

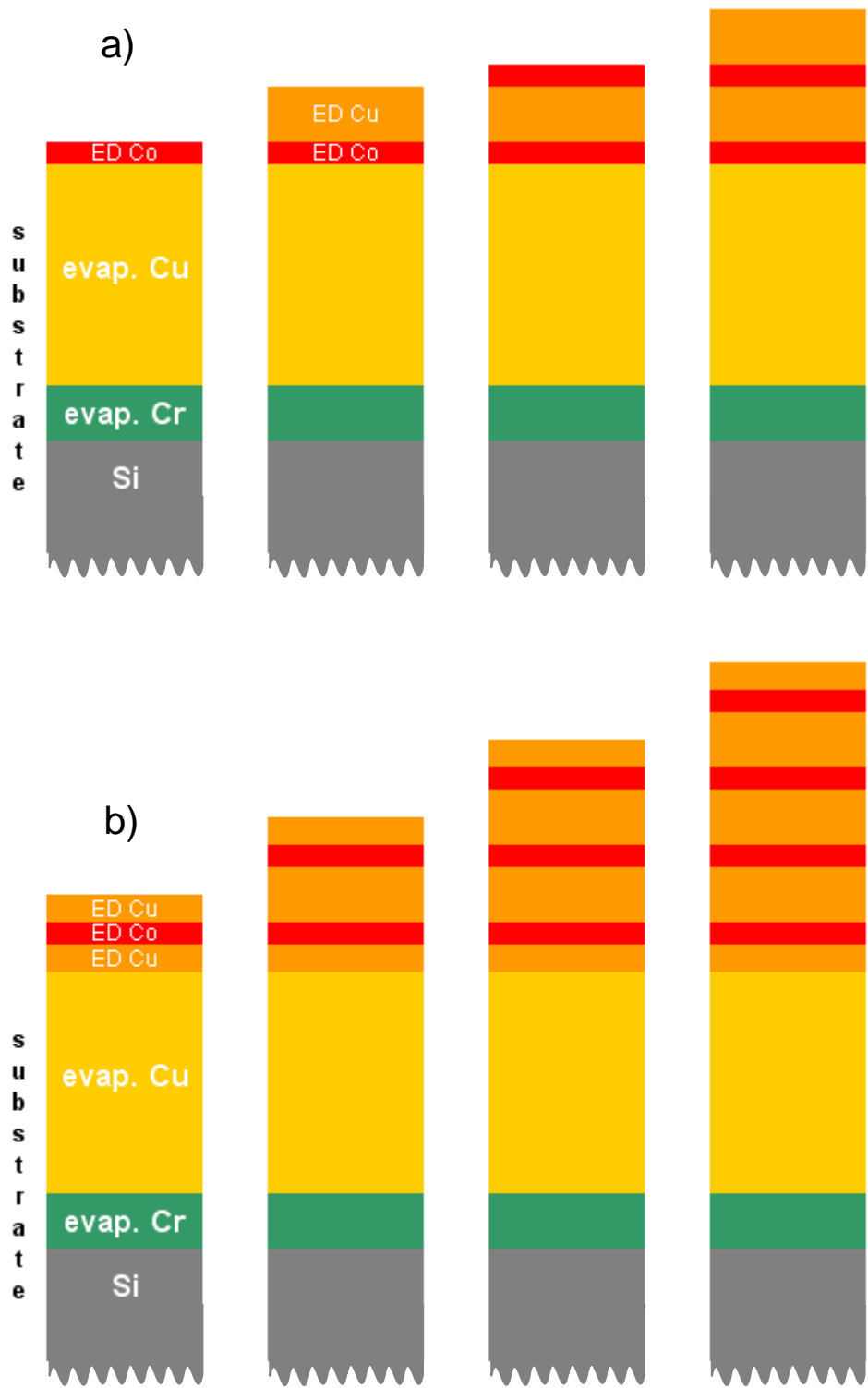

*Fig. 1* Schematic cross-sectional view of the investigated layer structures in (a) series 1 and (b) series 2.

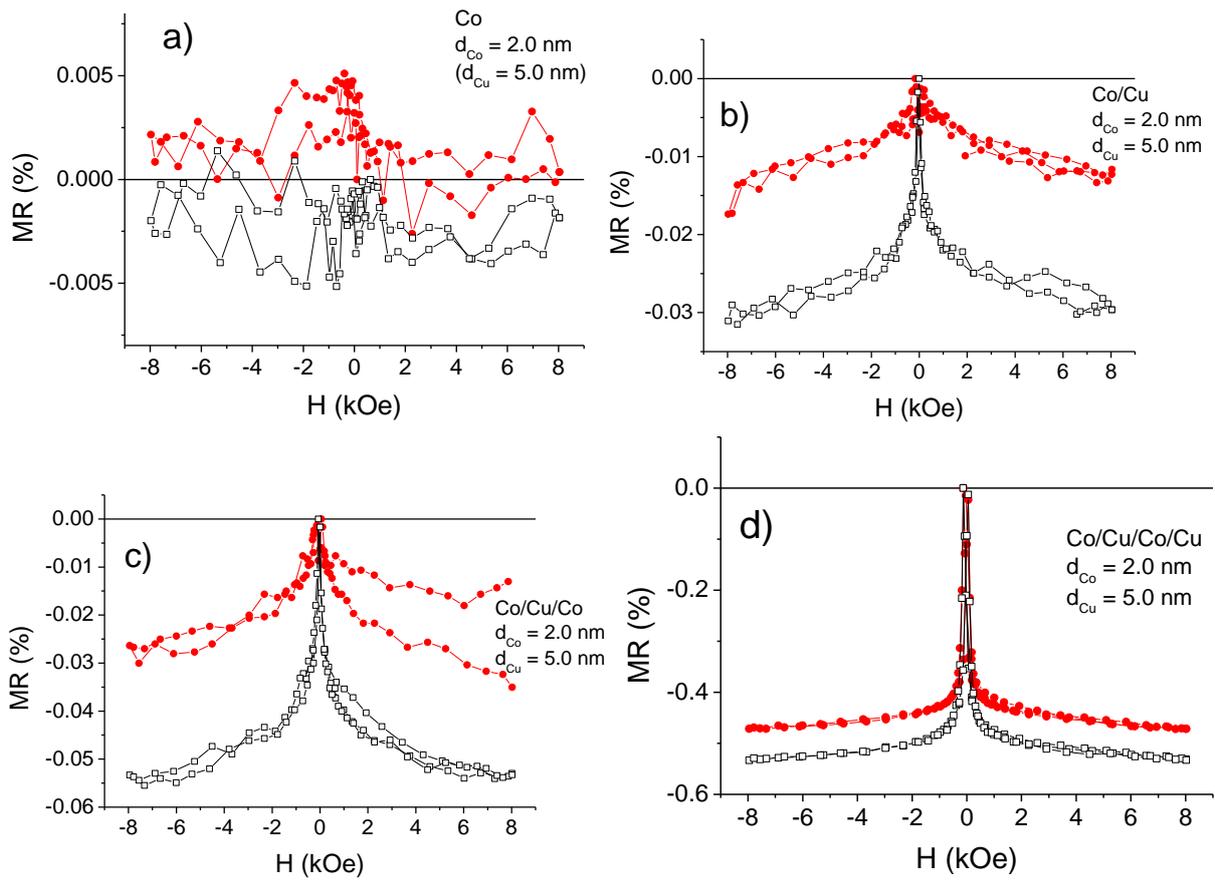

*Fig. 2* MR(*H*) curves for the layer stacks (a) Si/Cr/Cu//Co, (b) Si/Cr/Cu//Co/Cu, (c) Si/Cr/Cu//Co/Cu/Co and (d) Si/Cr/Cu//Co/Cu/Co/Cu of series 1. Circles: LMR data, squares: TMR data.

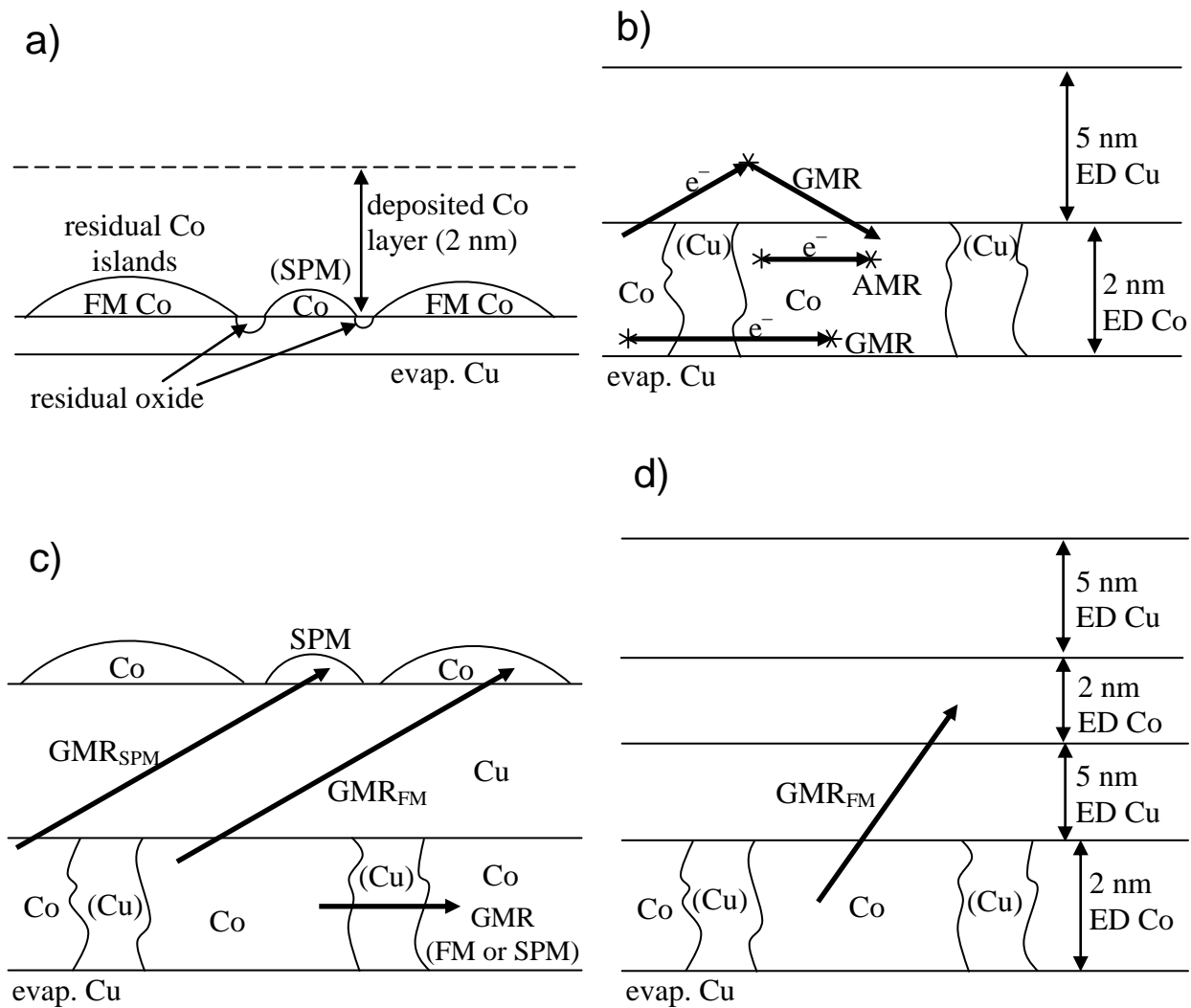

*Fig. 3* Sketches of the structure of the first four samples of series 1. The thicker/broader arrows mark the electron paths. Note that not all possible spin-dependent electron scattering paths are indicated in each sketch. The notation "(Cu)" indicates Cu-rich regions in the magnetic layer.

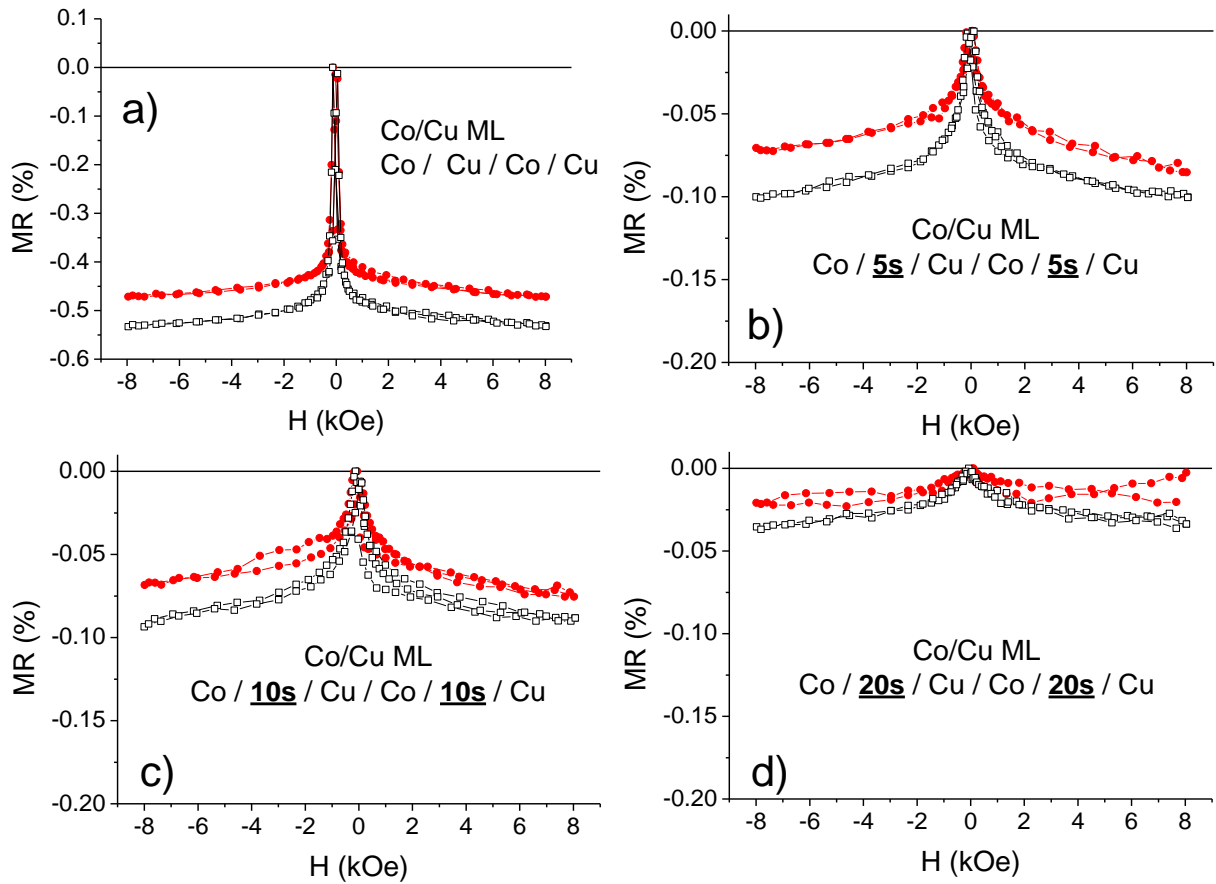

*Fig. 4* MR(*H*) curves for the multilayers Co/T/Cu/Co/T/Cu with (a) *T* = 0, (b) *T* = 5 s, (c) *T* = 10 s and (d) *T* = 20 s. The Co and Cu layer thicknesses were 2 nm and 5 nm, respectively. Circles: LMR data, squares: TMR data. Note the larger ordinate scale for (a) than for (b), (c) and (d). Note that the sample in (a) is identical with that in Fig. 2d.

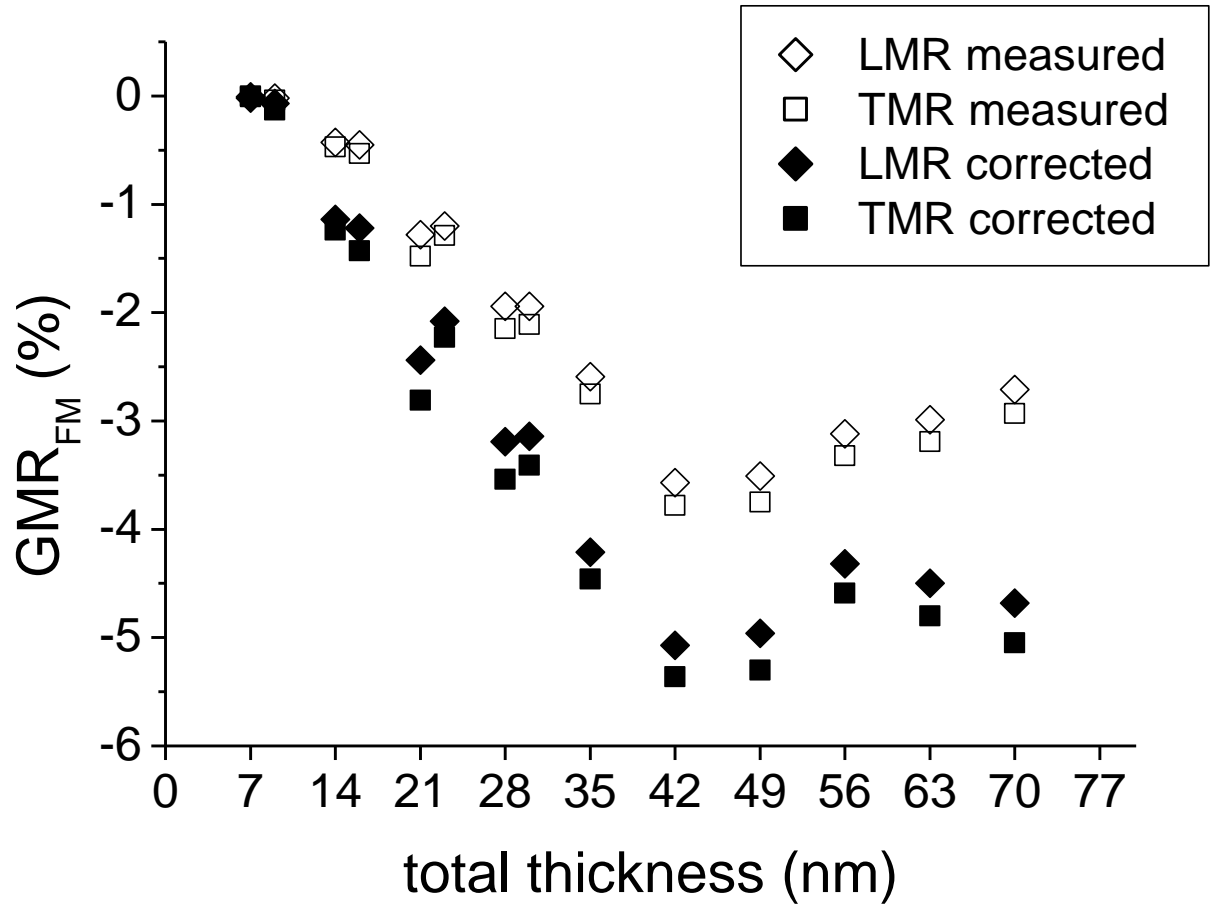

*Fig. 5*  Evolution of the FM contribution to the GMR with total deposit thickness for series 1 in which the multilayer is formed by the alternate deposition of Co(2 nm) magnetic and Cu(5 nm) non-magnetic layers. The open symbols are the measured data and the filled symbols are the GMR values corrected for the shunting effect of the substrate.

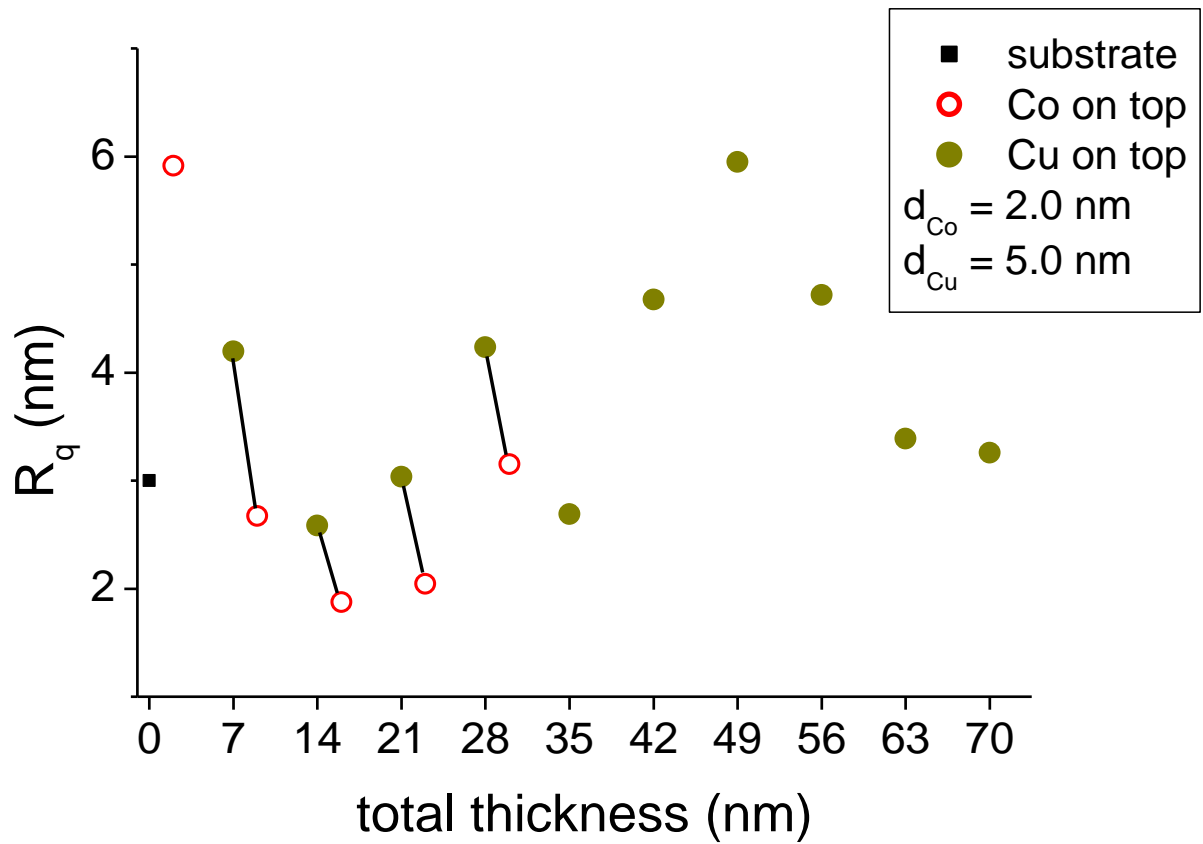

*Fig. 6* Evolution of the surface roughness parameter $R_q$ with total deposit thickness for series 1. The lines connecting the full and open circles are just to indicate the sequence "top Cu layer" → "top Co layer".

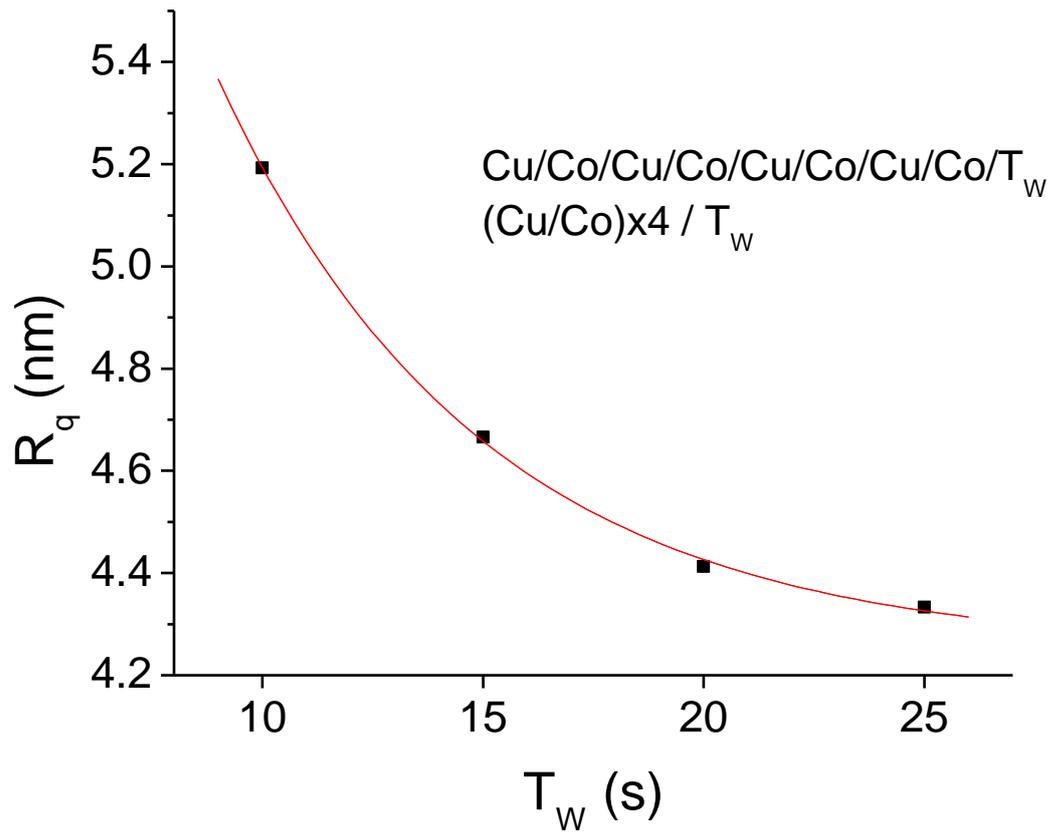

*Fig. 7* Evolution of the surface roughness parameter $R_q$ with the change of the waiting time ($T_W$).

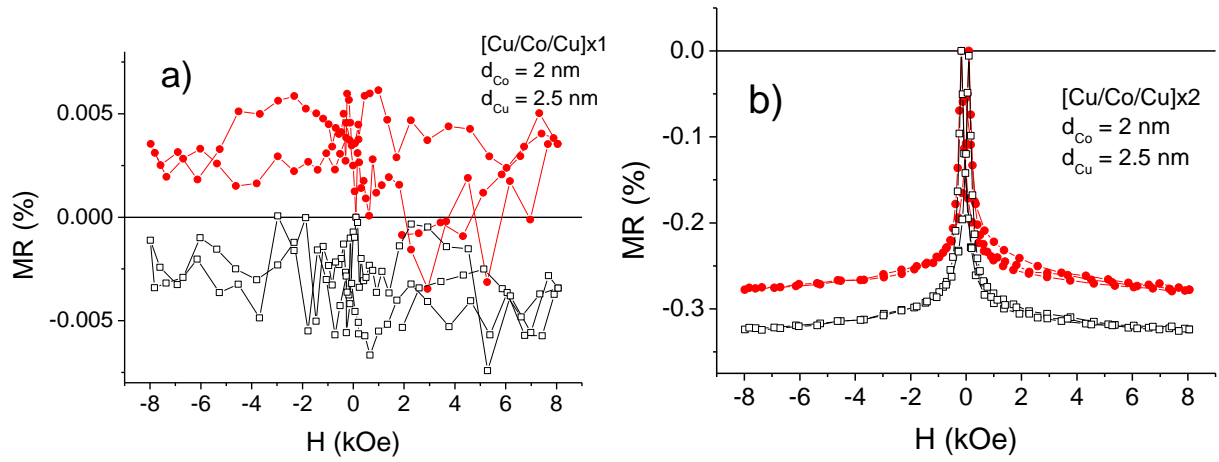

*Fig. 8* MR(*H*) curves for the layer stacks (a) Si/Cr/Cu//[Cu/Co/Cu]x1 and (b) Si/Cr/Cu//[Cu/Co/Cu]x2 of series 2. Circles: LMR data, squares: TMR data.

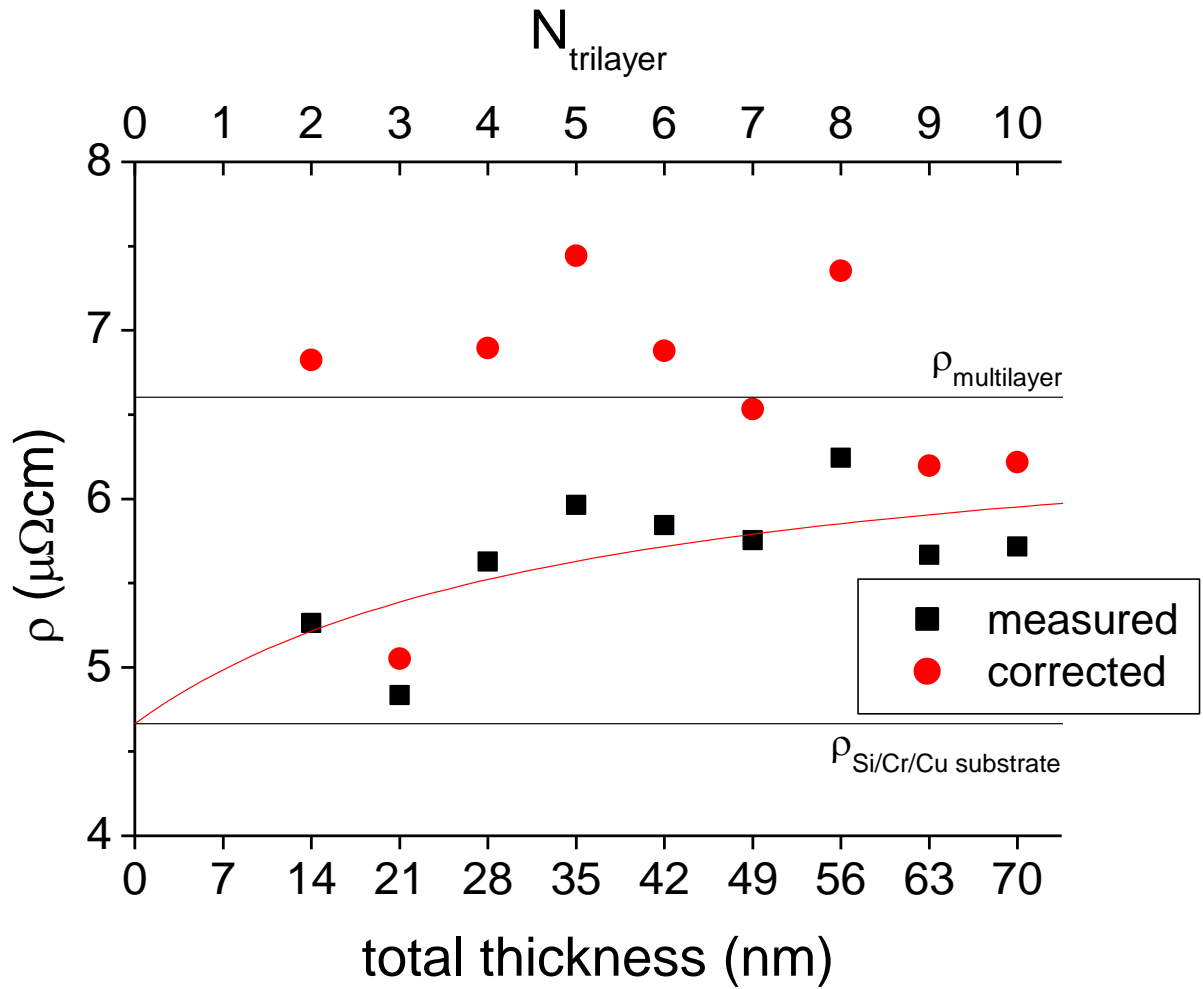

*Fig. 9* Evolution of the zero-field resistivity $\rho_0$ with total deposit thickness for series 2. The line through the measured data points is the result of a fit to eq. (1). The horizontal lines correspond to the resistivity of Si/Cr/Cu substrate and the average resistivity of the multilayers obtained from the fit.

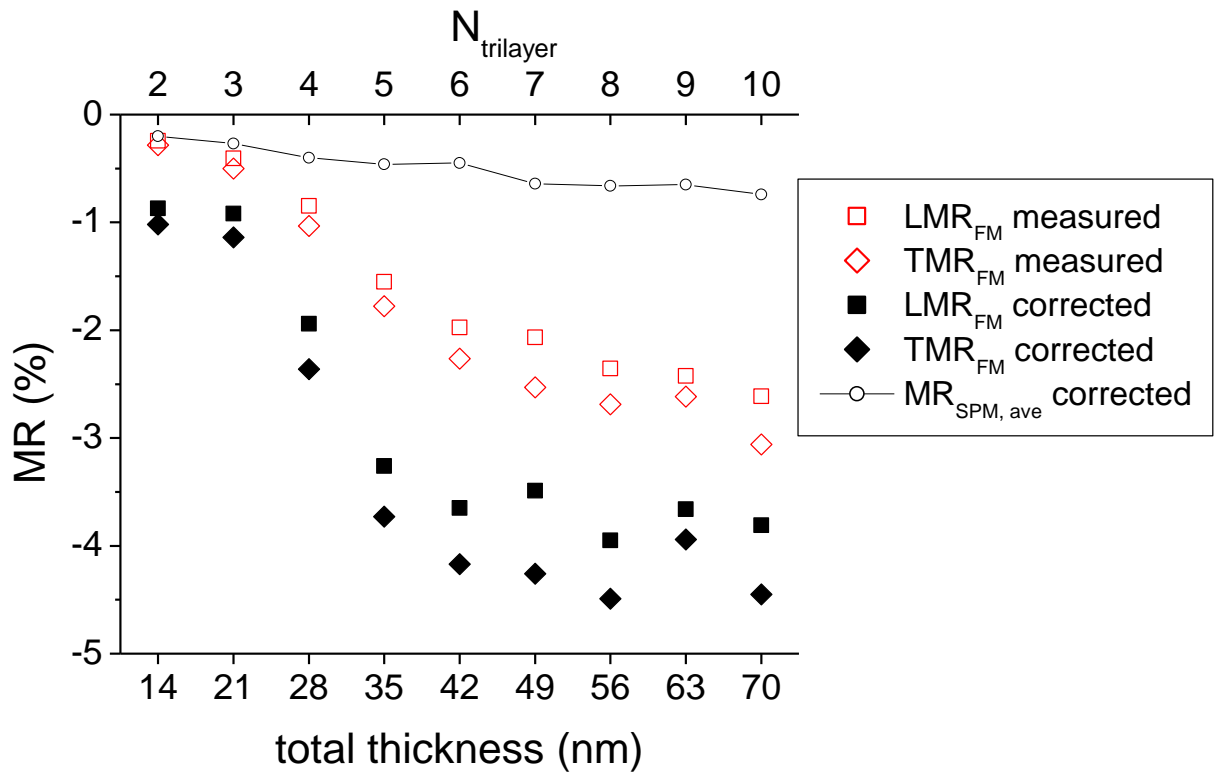

*Fig. 10* Evolution of the FM contribution to the GMR with total deposit thickness for series 2 in which the multilayers are built by the deposition of [Cu(2.5nm/Co(2nm)/Cu(2.5nm)] trilayer units. The open rectangles are the measured data and the filled rectangles are the GMR values corrected for the shunting effect of the substrate. The circles show the SPM-contribution averaged out for the longitudinal and the transverse component.

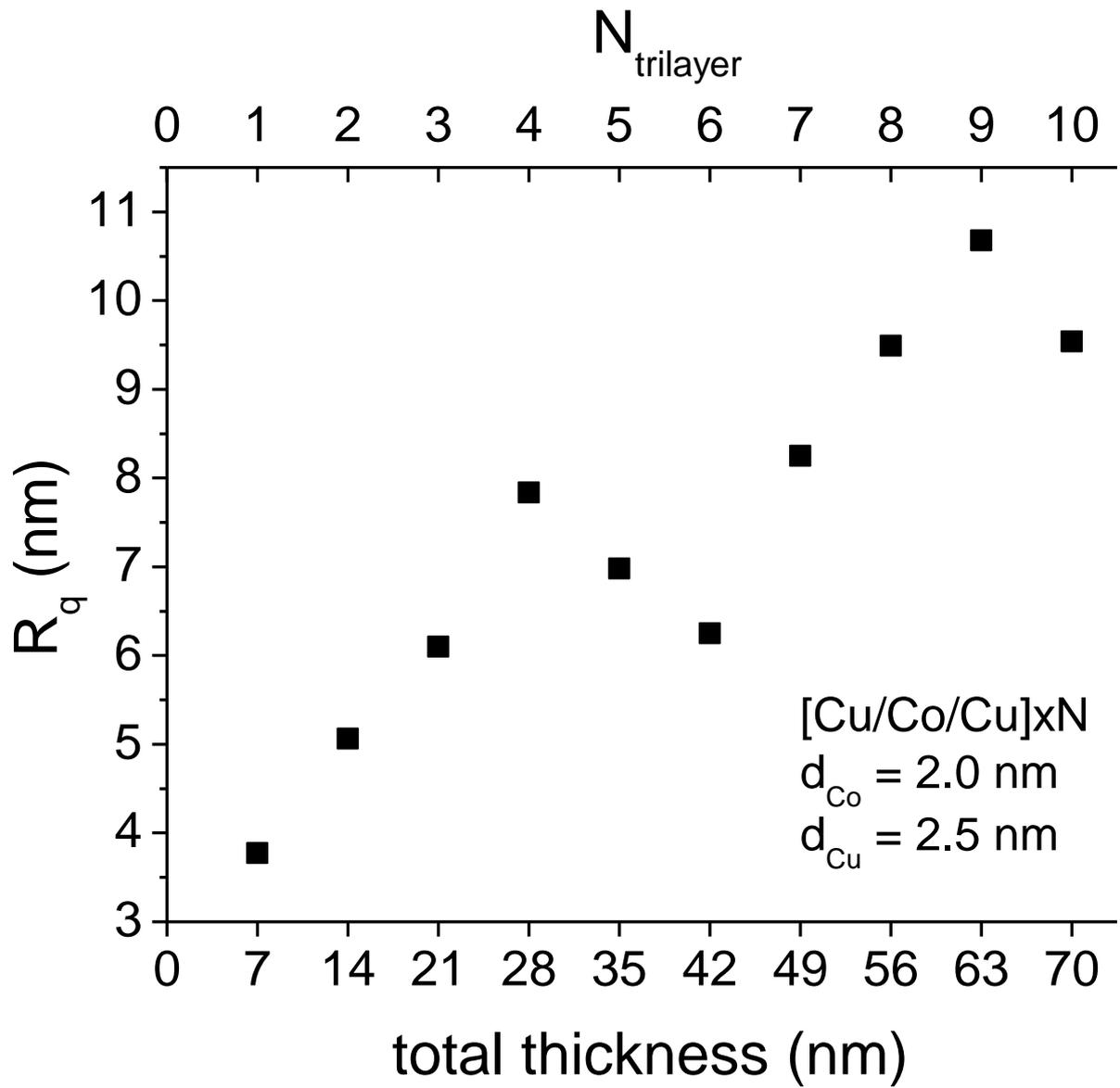

*Fig. 11* Evolution of the surface roughness parameter $R_q$ with total deposit thickness for series 2.

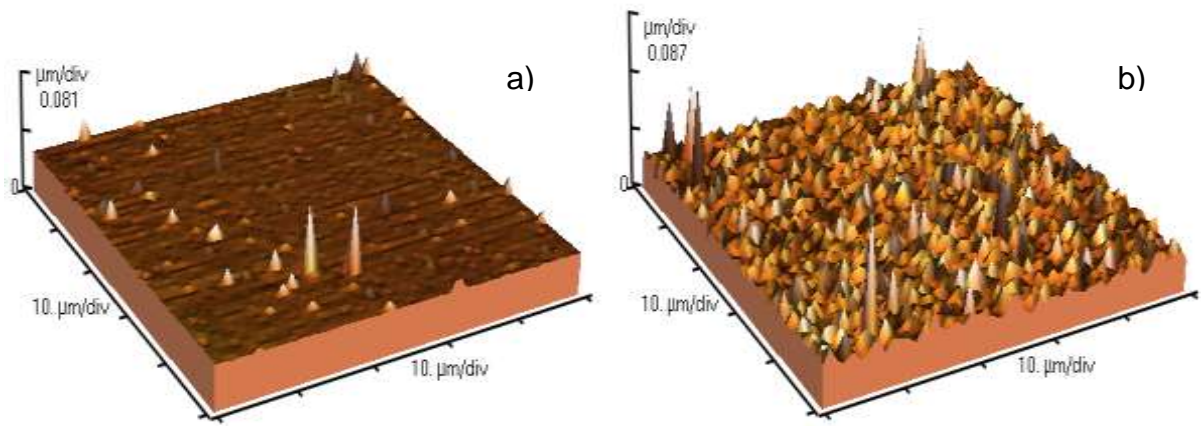

*Fig. 12* Representative AFM images of the surface of (a) the Si/Cr/Cu substrate and (b) the Co/Cu multilayer with 70 nm total thickness from series 2?. The images were obatined by smoothing the original records?. The rms roughness values are calculated by the data acquisition and evaluation computer of the AFM equipment from the measured height values of the whole image.